\documentclass[aps,prc,nofootinbib,onecolumn,groupedaddress,showpacs,showkeys]{revtex4}

\usepackage{graphicx}
\usepackage{epsfig,latexsym,amssymb}
\usepackage{amsmath}
\usepackage{bm}
\usepackage{slashed}
\usepackage{subfigure}
\usepackage{comment}
\usepackage{color}

\newcommand{\be}{\begin{equation}}
\newcommand{\ee}{\end{equation}}
\newcommand{\bea}{\begin{eqnarray}}
\newcommand{\eea}{\end{eqnarray}}

\newcommand{\as}{\alpha_s}

\def\eq#1{{Eq.~(\ref{#1})}}
\def\fig#1{{Fig.~\ref{#1}}}
\newcommand{\ben}{\begin{eqnarray*}}
\newcommand{\een}{\end{eqnarray*}}

\newcommand{\msbar}{\mu_{\overline {\text{MS}}}}
\newcommand{\Lmsbar}{\Lambda_{\overline {\text{MS}}}}

\begin{document}

\title{Particle multiplicities in the central region of high-energy
  collisions from $k_T$-factorization with running coupling corrections}
\author{Adrian Dumitru$^{1,2,3}$, Andre V. Giannini$^{4}$, Matthew Luzum$^{4}$ and Yasushi Nara$^5$}
\affiliation{$^1$ Department of Natural Sciences, Baruch College, CUNY,
17 Lexington Avenue, New York, NY 10010, USA \\
$^2$ The Graduate School and University Center, The City
  University of New York, 365 Fifth Avenue, New York, NY 10016, USA \\
$^3$ Physics Department, Brookhaven National
  Laboratory, Upton, NY 11973, USA \\
$^4$ Instituto de F\'{i}sica, Universidade de S\~ao Paulo, Rua do Mat\~ao
1371,  05508-090 S\~ao Paulo-SP, Brazil \\
$^5$ Akita International University, Yuwa, Akita-city 010-1292, Japan
}

\begin{abstract}
Horowitz and Kovchegov have derived a $k_T$-factorization formula for
particle production at small $x$ which includes running coupling
corrections. We perform a first numerical analysis to confront the
theory with data on the energy and centrality dependence of particle
multiplicities at midrapidity in high-energy p+A (and A+A)
collisions. Moreover, we point out a strikingly different dependence
of the multiplicity per participant on $N_\text{part}$ in p+Pb
vs.\ Pb+Pb collisions at LHC energies, and argue that the observed
behavior follows rather naturally from the convolution of
the gluon distributions of an asymmetric vs.\ symmetric projectile and target.
\end{abstract}

\maketitle
\vspace{1cm}

The Color Glass Condensate (CGC) approach to particle production in
high-energy collisions conjectures that the energy and system size
dependence of the $p_T$-integrated multiplicity can be computed in
weak coupling. The qualitative argument for this conjecture is that
the running coupling at the particle production vertex would be
effectively evaluated at a scale of order the semi-hard ``saturation
scale'' $Q_s\gg \Lambda_\text{QCD}$, even at low $p_T$.

McLerran and Venugopalan have shown that such a semi-hard scale indeed
emerges for a large nucleus due to the high density of valence color
charge per unit transverse area~\cite{MV}. Furthermore, the running
coupling Balitsky-Kovchegov (rcBK) equation~\cite{rcBK} for the
unintegrated gluon distribution (UGD) shows that the saturation scale grows
with energy. Most gluons ``in the wave function'' of a hadron or
nucleus have transverse momentum $k_T\sim Q_s$, suppressing the
sensitivity to the infrared, $k_T \sim \Lambda_{\rm
  QCD}$~\cite{Mueller:1999wm}; c.f.\ \fig{fig:phibar} below.

There have been many studies of the energy and centrality dependence
of particle production in p+A and A+A collisions within the
$k_T$-factorization approach~\cite{Gribov:1984tu}, using UGDs which
exhibit ``saturation'' at $k_T <Q_s$~\cite{Kharzeev:2001gp,
  Kharzeev:2004if, Levin:2010dw, Tribedy:2010ab, Albacete:2012xq}.
Whenever running of the coupling has been considered, an ad-hoc choice
for the scale of $\alpha_s(Q)$ in the $k_T$-factorization formula had
to be made\footnote{The issue of running coupling corrections also
  arises in fully numerical ``dense-dense''
  computations~\cite{Schenke:2012fw} which do not employ
  $k_T$-factorization.}. For example, ref~\cite{Albacete:2012xq}
assumed that the coupling is evaluated at $\text{max}\,
|\vec{p}_T\pm\vec{k}_T|/2$ so as to avoid the infrared regime (thanks
to $k_T\sim Q_s$, as mentioned above). While in practice the
sensitivity to such ad-hoc running coupling prescriptions may not be
very large, it is clearly worthwhile to assess running coupling
corrections from a more solid theoretical basis. Previous computations
of particle production in the central region relied on expressions
derived for fixed coupling, and running was implemented {\it a
  posteriori} by hand.

Horowitz and Kovchegov have derived a $k_T$-factorization formula
beyond LO to include running coupling
corrections~\cite{Horowitz:2010yg} (also see
ref.~\cite{KovchegovWeigert}) to single-inclusive (small-$x$) gluon
production in the scattering of two valence quarks. Their expression
results from a resummation of the relevant one-loop corrections into
the running of the coupling. They propose the following generalization to
hadron-hadron or hadron-nucleus collisions:
\begin{eqnarray}\label{eq:rcktfact}
  \frac{d^3 \sigma}{d^2 k \, dy} \, = \, N\,  \frac{2 \, C_F}{\pi^2} \,
  \frac{1}{{\bm k}^2} \, \int d^2q \int d^2b \, d^2b'\,
       {\overline \phi}_{h_1} ({\bm q}, y,{\bm b})
  \, {\overline \phi}_{h_2} ({\bm k} - {\bm q}, Y-y,{\bm b}-{\bm b'})
  \,
  \frac{\as \left(
      \Lambda_\text{coll}^2 \, e^{-5/3} \right)}{\as \left( Q^2 \,
      e^{-5/3} \right) \, \as \left( Q^{* \, 2}\, e^{-5/3} \right)} \,\,.
\end{eqnarray}
(Our notation follows ref.~\cite{Horowitz:2010yg}; ${\bm k}$ now
denotes the transverse momentum of the produced gluon while ${\bm q}$
and ${\bm k}-{\bm q}$ are the ``intrinsic'' transverse momenta from
the gluon distributions.)  This distribution of gluons in transverse
momentum and rapidity has to be convoluted with a fragmentation
function in order to obtain the $p_T$-distribution of produced
hadrons. \eq{eq:rcktfact} implicitly assumes that collinear
factorization applies in fragmentation.  $\Lambda_\text{coll}^2$ is a
collinear infrared cutoff which should match the scale of the
fragmentation function typically chosen as $\mu^2_\text{FF} \simeq
p_T^2$. We have computed hadron transverse momentum distributions in
p+A collisions in this way and shall report our findings
elsewhere. Here, we are primarily concerned with the $p_T$-integrated
multiplicity where the most relevant regime is that around the average
$p_T$. For this regime we employ a simple model fragmentation function
$D(z,\mu^2_\text{FF}) \sim \delta(1-z)$. For the observables
considered here slight modifications of this fragmentation function
mainly affect the normalization in \eq{eq:rcktfact} and can be
absorbed into $N$. The normalization also absorbs ``K-factors'' due to
higher order corrections and will be fixed by matching to data.

The unintegrated gluon distribution is given by
\begin{equation}\label{eq:rc_ktglueA}
  {\overline \phi} ({\bm k}, y,{\bm b})  =  \frac{C_F}{(2 \pi)^3} \,
  \int d^2 r \, e^{- i {\bm k} \cdot {\bm r}} \ \nabla^2_r \,
  \mathcal{N}_{A} ({\bm r}, y,{\bm b})\,.
\end{equation}
Note that these functions do not involve a factor of
$1/\alpha_s(k^2)$; instead, the factors of the inverse coupling with
the appropriate scale appear explicitly in \eq{eq:rcktfact}.
$\mathcal{N}_{A}({\bm r}, y,{\bm b})$ denotes the forward (adjoint)
dipole scattering amplitude at impact parameter ${\bm b}$. We assume a
uniform gluon density within a proton\footnote{For p+p collisions, not
  considered here, a more detailed model of the impact parameter
  dependence of $ \mathcal{N}_{A}$ is
  required~\cite{Levin:2010dw,Tribedy:2010ab}. Computing the impact
  parameter dependence of the gluon distribution of a proton directly
  from small-$x$ evolution is still an unresolved
  problem~\cite{GolecBiernat:2003ym}.}.  $\mathcal{N}_{A}(r)$
approaches a constant as $r\to\infty$ and so ${\overline \phi} ({\bm
  k})\sim k^2$ vanishes at low transverse momentum. For $k^2\gg
Q_s^2$, on the other hand, ${\overline \phi}(k)\sim 1/k^2$. In the
absence of the non-linear corrections to small-$x$ evolution present
in the BK equation this behavior would extend down to low $k_T$.

\begin{figure}[htb]
\begin{center}
\includegraphics[width=8.9cm]{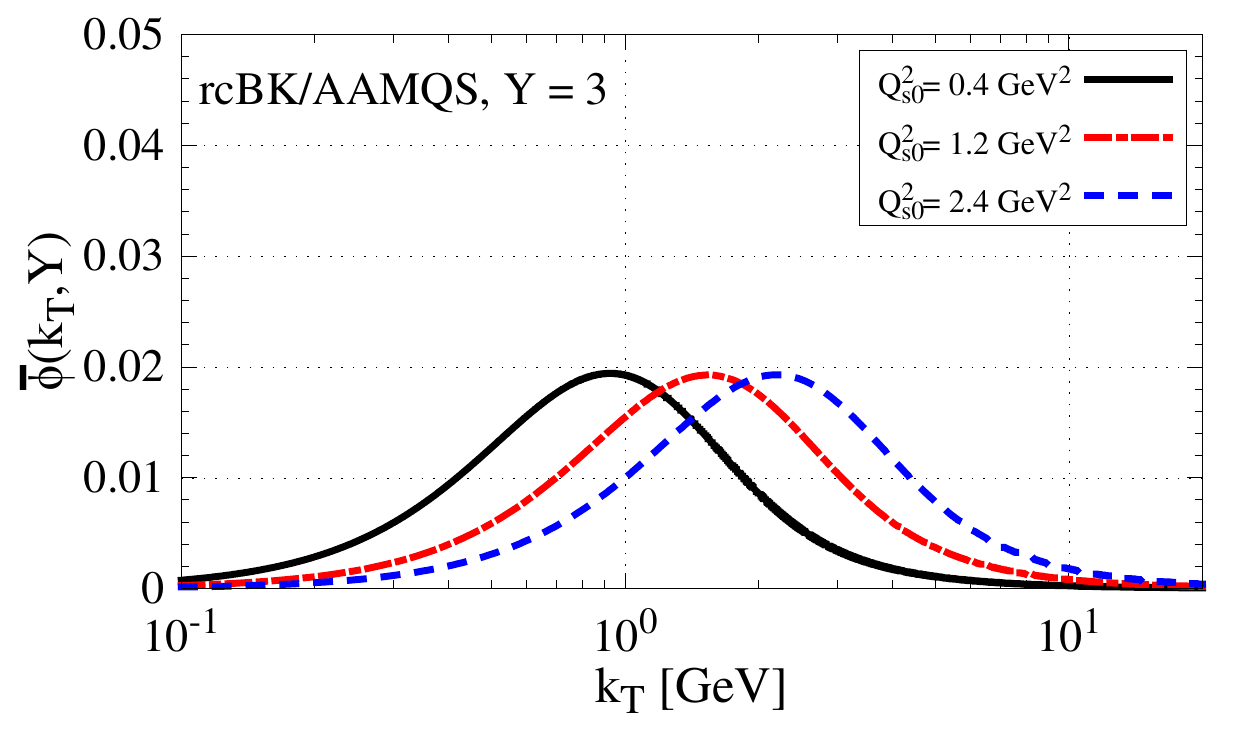}
\end{center}
\vspace*{-7mm}
\caption[a]{Impact parameter averaged, unbiased unintegrated gluon
  distribution ${\overline \phi}(k)$ at evolution rapidity $Y=3$ for a
  target with a thickness of one, three, and six nucleons,
  respectively. The peak of this function defines the saturation scale
$Q_s(Y)$. (The quoted values of $Q_{s0}^2$ refer to the
  saturation scale squared for an adjoint dipole at $x_0=0.01$.)}
\label{fig:phibar}
\end{figure}
The numerical form of $\mathcal{N}_{A}$ employed here is identical to
that used previously in ref.~\cite{Albacete:2012xq}; it has been
obtained in ref.~\cite{AAMQS} by solving the rcBK
equation\footnote{The BK equation actually provides the forward
  scattering amplitude for a fundamental dipole, averaged over
  configurations. At large $N_c$, from group theory, one obtains the
  average scattering amplitude for an adjoint dipole as
  $\mathcal{N}_{A} = 2\mathcal{N}_{F} - \mathcal{N}^2_{F} + {\cal
    O}(N_c^{-2})$.}. Here, we restrict to using their solution for MV
model initial condition since $p_T$-integrated observables for their
other UGD sets are not much different at small $x$. Finally, we stress
that the dipole forward scattering amplitude obtained in
ref.~\cite{Albacete:2012xq} has been averaged over all BK gluon
emissions without bias.  A plot of ${\overline \phi}(k)$ is shown in
\fig{fig:phibar}. This corresponds to the unintegrated gluon
distribution after 3 units of rcBK rapidity evolution.

The factors of the inverse coupling in \eq{eq:rcktfact} are determined
by the coefficient of the one-loop $\beta$-function (with
$N_c=N_f=3$), and we set $\Lmsbar=0.24$~GeV. The $Q^2$-dependence of
the coupling is given by
\begin{align} \label{eq:Qscale}
  \ln \frac{Q^2}{\msbar^2} \, & = \, \frac{1}{2} \, \ln \frac{{\bm
      q}^2 \, ({\bm k} - {\bm q})^2}{\msbar^4} - \frac{1}{4 \, {\bm
      q}^2 \, ({\bm k} - {\bm q})^2 \, \left[ ({\bm k} - {\bm q})^2 -
      {\bm q}^2 \right]^6} \, \Bigg\{ {\bm k}^2 \, \left[ ({\bm k} -
    {\bm q})^2 - {\bm q}^2 \right]^3 \notag \\[5pt]
  & \quad \; \times \, \bigg\{ \left[ \left[({\bm k}-{\bm
        q})^2\right]^2 - \left({\bm q}^2\right)^2 \right] \, \left[
    \left({\bm k}^2\right)^2 + \left(({\bm k}-{\bm q})^2 - {\bm
        q}^2\right)^2 \right] + 2 \, {\bm k}^2 \, \left[ \left({\bm
        q}^2\right)^3 - \left[({\bm k}-{\bm q})^2\right]^3
  \right] \notag \\[5pt]
  & \quad \; - {\bm q}^2 \, ({\bm k} - {\bm q})^2 \left[ 2 \,
    \left({\bm k}^2\right)^2 + 3 \, \left[({\bm k} - {\bm q})^2 - {\bm
        q}^2\right]^2 - 3 \, {\bm k}^2 \, \left[({\bm k} - {\bm q})^2
      + {\bm q}^2\right] \right] \, \ln
  \left( \frac{({\bm k} - {\bm q})^2}{{\bm q}^2} \right) \bigg\} \notag \\[5pt]
  & + \, i \, \left[ ({\bm k} - {\bm q})^2 - {\bm q}^2 \right]^3 \, \,
  \bigg\{ {\bm k}^2 \, \left[ ({\bm k} - {\bm q})^2 - {\bm q}^2\right]
  \, \left[ {\bm k}^2 \, \left[ ({\bm k} - {\bm q})^2 + {\bm
        q}^2\right] - \left({\bm q}^2\right)^2 -
    \left[({\bm k}-{\bm q})^2\right]^2 \right] \notag \\[5pt]
  & \quad \; + {\bm q}^2 \, ({\bm k} - {\bm q})^2 \, \left( {\bm k}^2
    \, \left[ ({\bm k} - {\bm q})^2 + {\bm q}^2\right] - 2 \,
    \left({\bm k}^2\right)^2 - 2 \, \left[ ({\bm k} - {\bm q})^2 -
      {\bm q}^2\right]^2 \right) \, \ln \left( \frac{({\bm k} - {\bm
        q})^2}{{\bm q}^2} \right) \bigg\} \notag \\[5pt]
  & \quad \; \times \, \sqrt{2 \, {\bm q}^2 \, ({\bm k} - {\bm q})^2 +
    2 \, {\bm k}^2 \, ({\bm k} - {\bm q})^2 + 2 \, {\bm q}^2 \, {\bm
      k}^2 - \left({\bm k}^2\right)^2 - \left({\bm q}^2\right)^2 -
    \left[({\bm k}-{\bm q})^2\right]^2} \Bigg\}\,\,,
\end{align}
$\ln \frac{Q^{*\, 2}}{\msbar^2}$ is given by the complex conjugate of
this expression so that the product $\ln \frac{Q^2}{\msbar^2} \ln
\frac{Q^{*\, 2}}{\msbar^2}$ is real, as it should be. In the limit
$q \ll k$ this simplifies to\footnote{The r.h.s.\ of
  \eq{eq:lnQ2_q2->0} differs from the expression given in eq.~(3.33)
  of ref.~\cite{Horowitz:2010yg}; the correct result, which we
  verified independently, was first communicated to us in private by
  Yu.~Kovchegov.}
\bea
\ln \frac{Q^2}{\msbar^2}\Bigr|_{q\to0} &=&
\ln \frac{k^2}{\msbar^2} + \frac{1}{2} - \frac{({\bm k}\cdot {\bm
q})^2}{k^2 q^2} - i \frac{{\bm k}\cdot {\bm
q}}{k^2 q^2} \sqrt{k^2 q^2 - ({\bm k}\cdot {\bm
    q})^2} \label{eq:lnQ2_q2->0}~,
\eea
so that
\be
\int \frac{d\phi_q}{2\pi} \left[
  \ln \frac{Q^2}{\msbar^2} \ln
  \frac{Q^{*\,2}}{\msbar^2}\right]_{q\to0}
= \ln^2 \frac{k^2}{\msbar^2} + \frac{1}{4}~.
\label{eq:lnQ2lnQ*2_q2->0}
\ee
Therefore, at high transverse momentum, and choosing the collinear
cutoff scale $\Lambda^2_\text{coll}=k^2$, the spectrum of produced gluons
is proportional to $\alpha_s(k^2)\, k^{-4} \ln^2 k^2$.

For $k\ll q$ we have
\bea
\ln \frac{Q^2}{\msbar^2}\Bigr|_{k\to0} &=& \ln\frac{q^2}{\msbar^2}~,
\label{eq:lnQ2_k->0}
\eea
Here, the dominant contribution to \eq{eq:rcktfact} is from $q\sim
Q_s$ since ${\overline \phi}(q)$ quickly decays when $q$ is far from
$Q_s$.  \eq{eq:lnQ2_k->0} shows that the distribution of produced
gluons is well defined at low transverse momentum, $k\to\Lmsbar$.  The
spectrum can be integrated over $k>\Lmsbar$ without encountering a
divergence. (It is not sensible to address the spectrum of ``gluons''
with $k<\Lmsbar$.)

Physically, $d\sigma/d^2kdy\sim \alpha_s(\Lambda^2_\text{coll}\sim
k^2)/k^2$ should rather level off when both collision partners are
dense; $k_T$-factorization fails here. For proton-nucleus collisions
with $Q_{s,A} \gg Q_{s,p}$ the actual contribution to $dN/dy$ from
$k<Q_{s,p}$ is small and we can therefore simply apply
\eq{eq:rcktfact} down to $k=\Lmsbar$.  For nucleus-nucleus collisions
this is not justified since there are many particles at
$\Lmsbar<k<Q_s$. Here, the correct spectrum below $Q_s$ can only be
obtained from a ``dense-dense'' computation which does not rely on
$k_T$-factorization. (To date, such calculations, for example,
refs.~\cite{Schenke:2012fw,Blaizot:2010kh}, have been performed only
for fixed coupling, or with ad-hoc running.) On the other hand,
phenomenological applications of $k_T$-factorization have been rather
successful in reproducing the dependence of the multiplicity in A+A
collisions on energy and
centrality~\cite{Kharzeev:2001gp,Tribedy:2010ab,Albacete:2012xq}. This
is presumably due to the fact that for large nuclei and high energies
this dependence is entirely determined by the single scale $Q_s$. In
any case, given those previous applications of $k_T$-factorization
with ad-hoc scale choice to the centrality dependence of the
multiplicity in A+A collisions it is certainly interesting to also see
the result obtained from \eq{eq:rcktfact}. Hence, we integrate
\eq{eq:rcktfact} from $k_\text{min}$ of order $\Lmsbar$. We have
checked the dependence of $dN/d\eta$ in Pb+Pb collisions at LHC energies
on $N_\text{part}$ for $k_\text{min}=\Lmsbar \dots 2\Lmsbar$
and obtained virtually identical curves.

\begin{figure}[htb]
\begin{center}
\includegraphics[width=8.9cm]{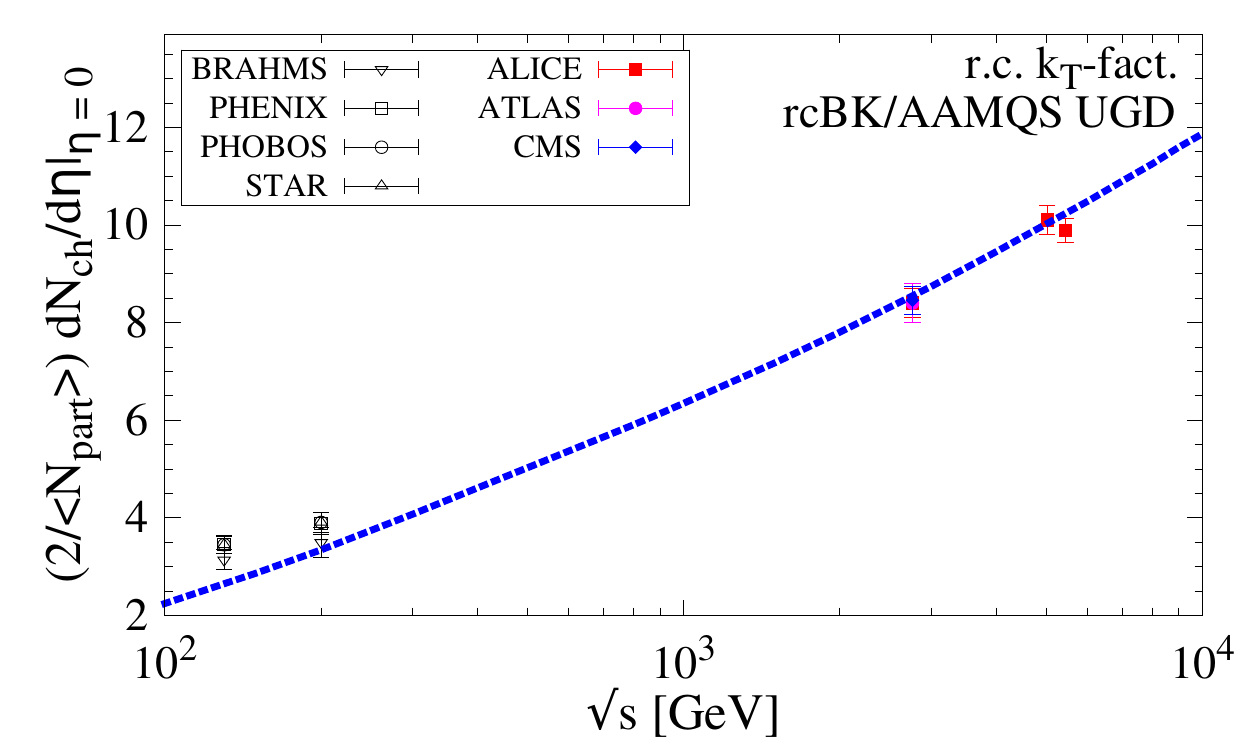}
\includegraphics[width=8.9cm]{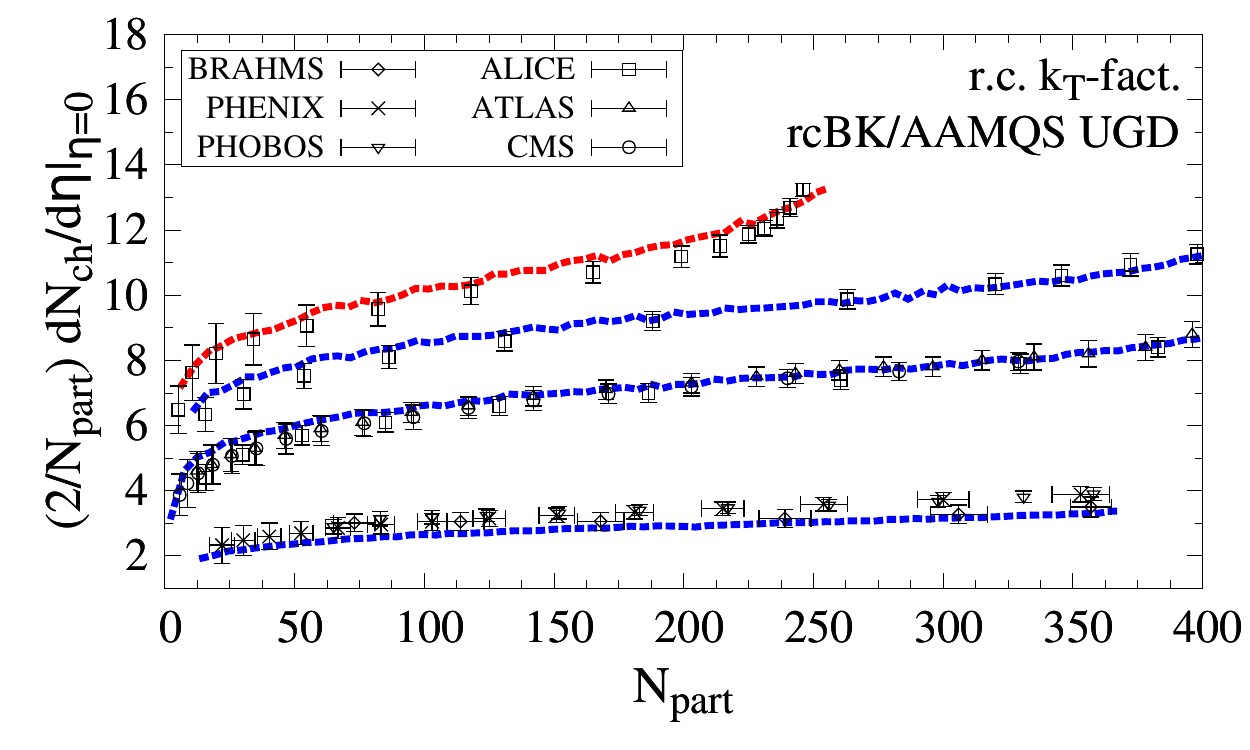}
\end{center}
\vspace*{-7mm}
\caption[a]{Left: Energy dependence of the multiplicity per
  participant pair in central (0-6\%) Xe+Xe/Au+Au/Pb+Pb collisions at
  $\sqrt{s}=200$~GeV, 2.76~TeV and 5.02~TeV, respectively.  Right:
  same as a function of $N_\text{part}$.  The curve and data points
  for 5.02~TeV have been scaled by 1.1 to improve visibility;
  moreover, our prediction and the new data for Xe+Xe at
  $\sqrt{s}=5.44$~TeV have been scaled by 1.25 in order to not overlap
  with the curves for Pb+Pb collisions.}
\label{fig:AA_Npart}
\end{figure}
To compute produced particle multiplicities in p+A and A+A collisions
we convolute \eq{eq:rcktfact} with a Monte-Carlo Glauber simulation,
which has been described in more detail in
refs.~\cite{Albacete:2012xq}. This allows us to compute the dependence
of the multiplicity on the number of participants. First results using \eq{eq:rcktfact}
were obtained in ref.~\cite{Duraes:2016yyg} for
minimum bias collisions with KLN
model~\cite{Kharzeev:2001gp,Kharzeev:2004if} gluon distributions.

\fig{fig:AA_Npart} shows our results for the multiplicity per
participant pair in A+A collisions at RHIC and LHC energies. We have
fixed the normalization factor in \eq{eq:rcktfact} to match to central
Pb+Pb collisions at 2.76~TeV; the same normalization has been used for
all other energies, centralities, and collision systems. The data
shown in \fig{fig:AA_Npart} is from
refs~\cite{Bearden:2001xw,Adler:2004zn, Back:2002uc, Abelev:2008ab,
  Chatrchyan:2011pb, ATLAS:2011ag, Aamodt:2010cz}. Our curves are very
close to those published in ref.~\cite{Albacete:2012xq} using ad-hoc
scale setting. The data at $\sqrt{s}=200$~GeV mainly probes the MV
model gluon distribution at the initial $x_0=0.01$ rather than
small-$x$ rcBK evolution. The multiplicity as a function of
$N_\text{part}$ shows the well known increase of $dN/d\eta$ per
participant towards more central collisions. It is driven by the
increasing overlap in transverse coordinate space of the 2d
projections of the nuclear Woods-Saxon distributions. This leads to
increasingly {\em symmetric} collision partners at any given point in
the transverse plane so that the convolution integral of the gluon
distributions in \eq{eq:rcktfact} increases as both transverse
momentum arguments can be near the ``saturation peak''.

We also show our prediction for Xe+Xe collisions
at 5.44~TeV in \fig{fig:AA_Npart}. We have updated the figure to
include new data for Xe+Xe collisions released by the ALICE
collaboration~\cite{Acharya:2018hhy}.

\begin{figure}[htb]
\begin{center}
\includegraphics[width=8.9cm]{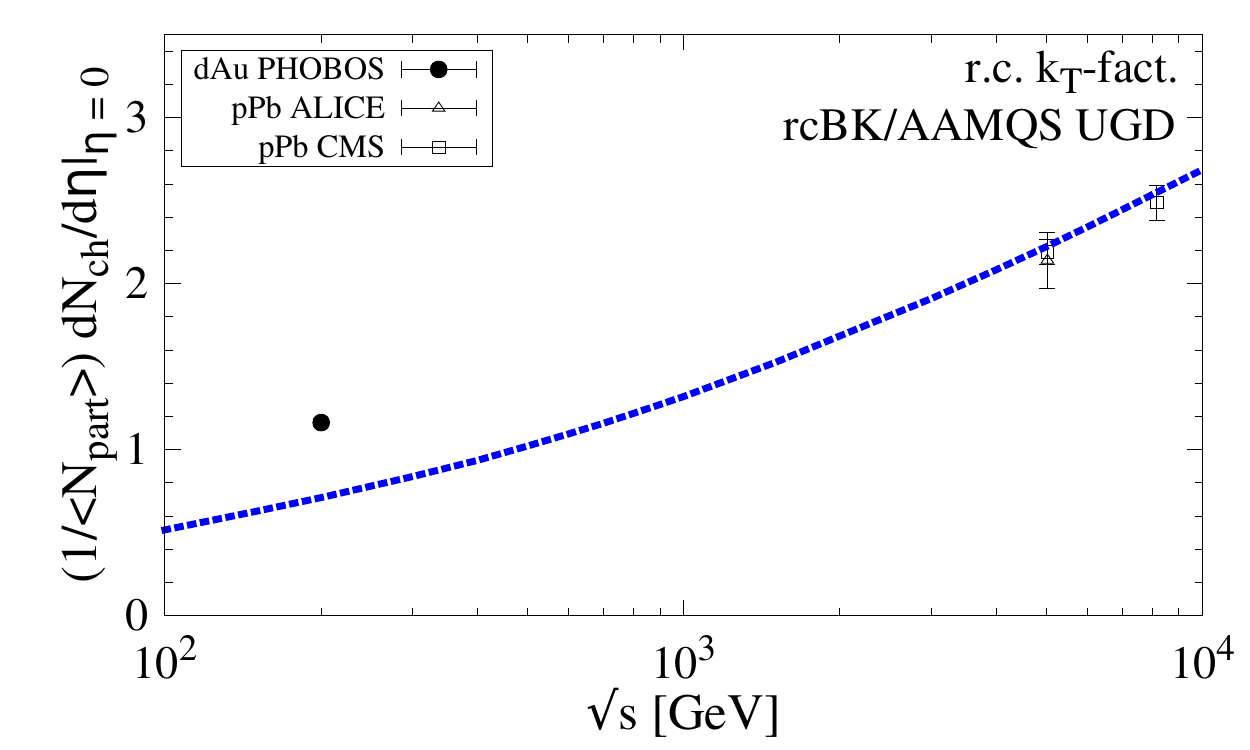}
\includegraphics[width=8.9cm]{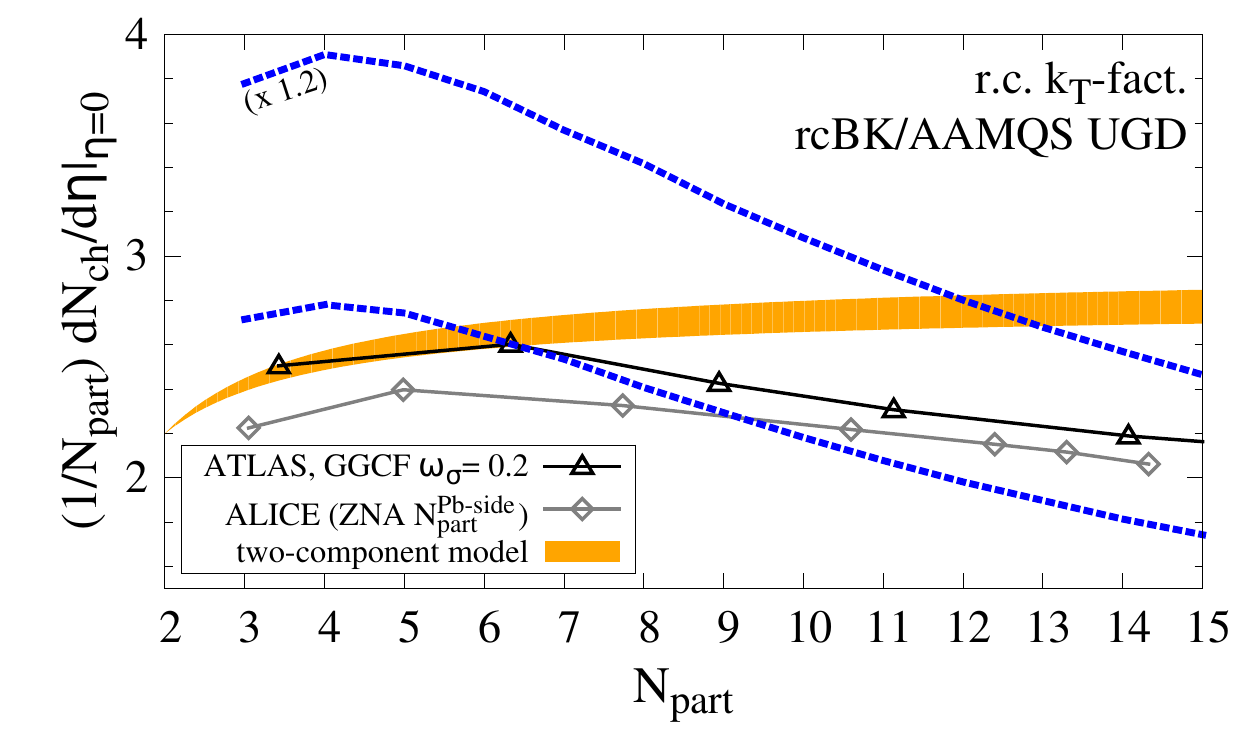}
\end{center}
\vspace*{-7mm}
\caption[a]{Energy and $N_\text{part}$ dependence of the multiplicity
  in p+Pb collisions at $\eta=0$. In the figure on the right the two
  short dashed lines correspond to our results for 5.02~TeV and
  8.16~TeV, respectively; the latter have been rescaled by a factor of
  1.2 for better visibility. The band corresponds to the ``soft+hard''
  two-component model in \eq{eq:soft+hard_pA}.}
\label{fig:pA_roots_Npart}
\end{figure}
\fig{fig:pA_roots_Npart} shows the multiplicity in p+A collisions as a
function of energy and of $N_\text{part}$; midrapidity ($\eta=0$)
corresponds to the CM frame. The data points are from
refs.~\cite{Back:2003hx,ALICE:2012xs,Sirunyan:2017vpr}.  The energy
dependence of the multiplicity obtained from the
r.c.\ $k_T$-factorization formula with rcBK UGDs compares well to the
measurements at LHC energies. The extrapolation to RHIC energy of
$\sqrt{s}=200$~GeV, however, is significantly too low. This is not
unexpected since at such energies one is sensitive mainly to the MV
model initial condition imposed at $x_0=0.01$ rather than to small-$x$
evolution. This should in fact fail, in particular for small systems,
since the MV model assumes a large nucleus. Improving results for
p/d+A collisions at RHIC energy (and $\eta\sim0$) will require an improved
theoretical understanding of the unintegrated gluon distribution
of a proton at $x=0.01$ and greater, as well as possibly additional
corrections to \eq{eq:rcktfact}.

The dependence of the multiplicity at LHC energies on $N_\text{part}$
is rather interesting.  Somewhat surprisingly perhaps, we find that
beyond $N_\text{part}\simeq4$ the multiplicity {\em per participant}
decreases slightly with $N_\text{part}$. This is due to the fact that
for increasingly asymmetric collisions the convolution in transverse
momentum space of the gluon distributions does not increase in
proportion to $N_\text{part}$. Simple considerations suggest that it
grows logarithmically (also see the discussion in
refs.~\cite{Kharzeev:2004if}). A numerical fit to the 5.02~TeV curve
shown in \fig{fig:pA_roots_Npart}(right), for $15\ge N_\text{part}\ge
5$, gives $\sim \ln^{1.25}(N_\text{part}) / N_\text{part}$.  In
contrast, A+A collisions become more symmetric as the impact parameter
decreases and the multiplicity per participant increases with
$N_\text{part}$.

Such a feature is also seen in data, as shown in
\fig{fig:pA_roots_Npart} (right), where we show
ALICE~\cite{Adam:2014qja} and ATLAS~\cite{Aad:2015zza} data for
$(1/N_\text{part})\, dN_\text{ch}/d\eta$ vs.\ $N_\text{part}$ in p+Pb
collisions at 5~TeV. In fact, in ref.~\cite{ALICE:2012xs} the ALICE
collaboration already noted that the multiplicity per participant in
NSD p+Pb collisions at 5~TeV (averaged over $N_\text{part}$~!) is 16\%
{\em lower} than in NSD p+p collisions interpolated to the same
collision energy. Remarkably, this trend appears to continue beyond
$N_\text{part}^\text{mb}\simeq 8$.

While the distribution of multiplicity is a quantity that can be
measured in a fairly direct way, the number of participants is not,
and in principle depends on the method of centrality selection.  We
refer to the above-mentioned publications for more detailed
discussions of the experimental centrality selections and their
determination of $N_\text{part}$, but here show the ALICE ``ZNA
$N_\text{part}^\text{Pb-side}$'' and the ATLAS ``GGCF
$\omega_\sigma=0.2$'' results~\cite{Aad:2015zza,Guzey:2005tk}, which are believed to be less model
dependent and may be the most suitable to compare to $N_\text{part}$
as used in our model.

These data exhibit a trend similar to the calculation, but with
somewhat flatter dependence on $N_\text{part}$.  This could be due to
the lack of a realistic impact parameter dependence of the proton-UGD
in our computations, and due to a bias on the gluon distribution
introduced by the experimental centrality selection. A more accurate
matching of $(1/N_\text{part})\, dN_\text{ch}/d\eta$ to the
measurements would entail accounting for such bias on the
configurations of small-$x$ gluon fields through
reweighting~\cite{Dumitru:2018iko}.

\begin{figure}[htb]
\begin{center}
\includegraphics[width=8.9cm]{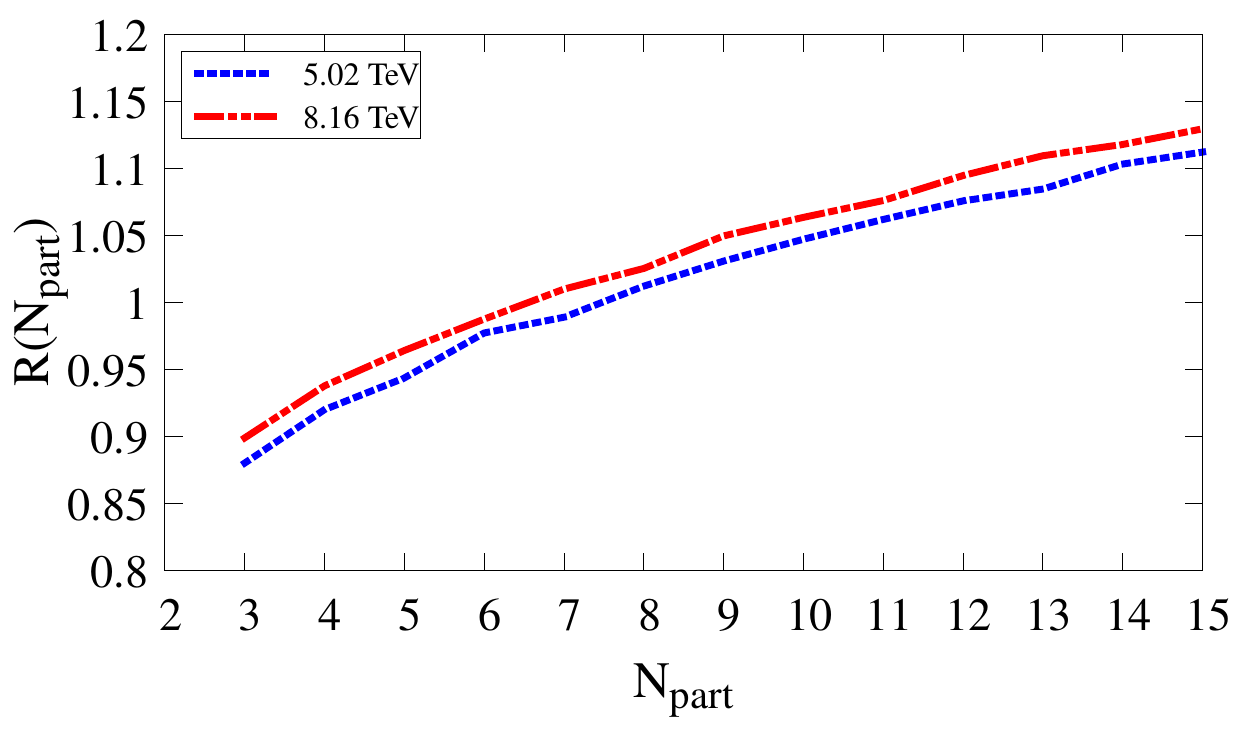}
\includegraphics[width=8.9cm]{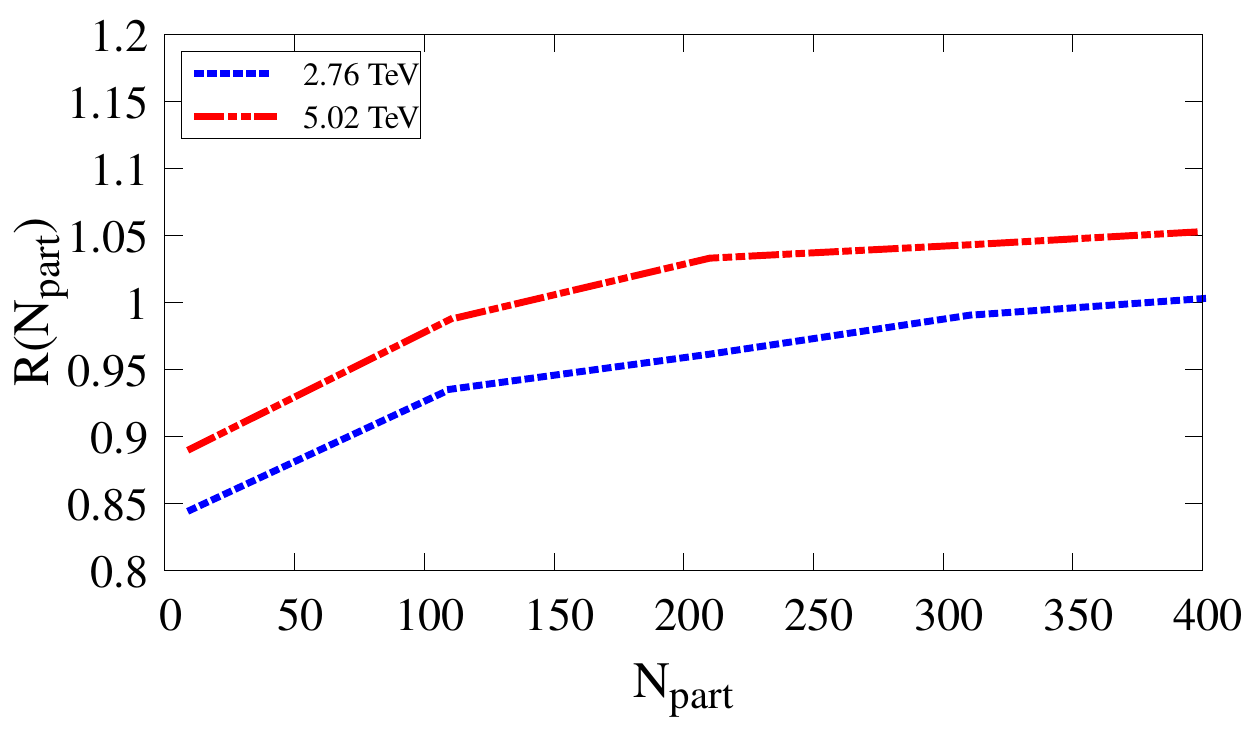}
\end{center}
\vspace*{-7mm}
\caption[a]{Ratios of charged particle multiplicities in p+Pb (left)
  and Pb+Pb (right) collisions obtained from a fixed-coupling
  $k_T$-factorization formula with running of $\alpha_s$ implemented
  by hand (see text) divided by eq.~(\ref{eq:rcktfact}).}
\label{fig:ratio_Nch}
\end{figure}
It is interesting to compare the prediction from \eq{eq:rcktfact} to a
$k_T$-factorization formula derived at fixed coupling, with running of
$\as$ implemented by hand, and with rcBK gluon distributions. The
latter corresponds to replacing in eq.~(\ref{eq:rcktfact}):
\bea
\as\left(\Lambda_\text{coll}^2 \, e^{-5/3} \right)
&\to&
\as\left(k^2\right) ~, \\
\as\left( Q^2 \, e^{-5/3} \right)
&\to&
\as\left(q^2\right) ~, \label{eq:alpha_q}\\
\as\left( Q^{* \, 2}\, e^{-5/3} \right)
&\to&
\as\left(({\bm k}-{\bm q})^2\right)  \label{eq:alpha_k-q} ~.
\eea
(Other prescriptions for running of $\as$ ``by hand'' exist, as
already mentioned in the introduction.) The
replacements~(\ref{eq:alpha_q},\ref{eq:alpha_k-q}), in particular, are
inspired by the fact that at fixed coupling the unintegrated gluon
distributions about $Q_s$ are of order $1/\alpha_s$.  Note that in the
${\bm q}\to0$ limit at high ${\bm k}$ the ratio of coupling constants
in \eq{eq:rcktfact} now approaches a ${\bm k}$-independent constant as
opposed to Eqs.~(\ref{eq:lnQ2_q2->0},\ref{eq:lnQ2lnQ*2_q2->0}). To
obtain the particle multiplicity we integrated over ${\bm k}$ from
$k_\text{min}=0.25$~GeV. Also, we again adjusted the normalization factor
$N_\text{f.c.}$ to the multiplicity in central Pb+Pb collisions at
2.76~TeV.

The resulting ratio is shown in \fig{fig:ratio_Nch}. Overall, the
f.c.\ formula with ad-hoc running of the coupling provides a fairly
satisfactory description of the dependence of the multiplicity on
$N_\text{part}$, so that the discrepancy to the r.c.\ formula is
fairly moderate. However, a systematically steeper rise of
$dN_\text{ch} / d\eta$ with $N_\text{part}$ is clearly visible.

\begin{figure}[htb]
\begin{center}
\includegraphics[width=8.9cm]{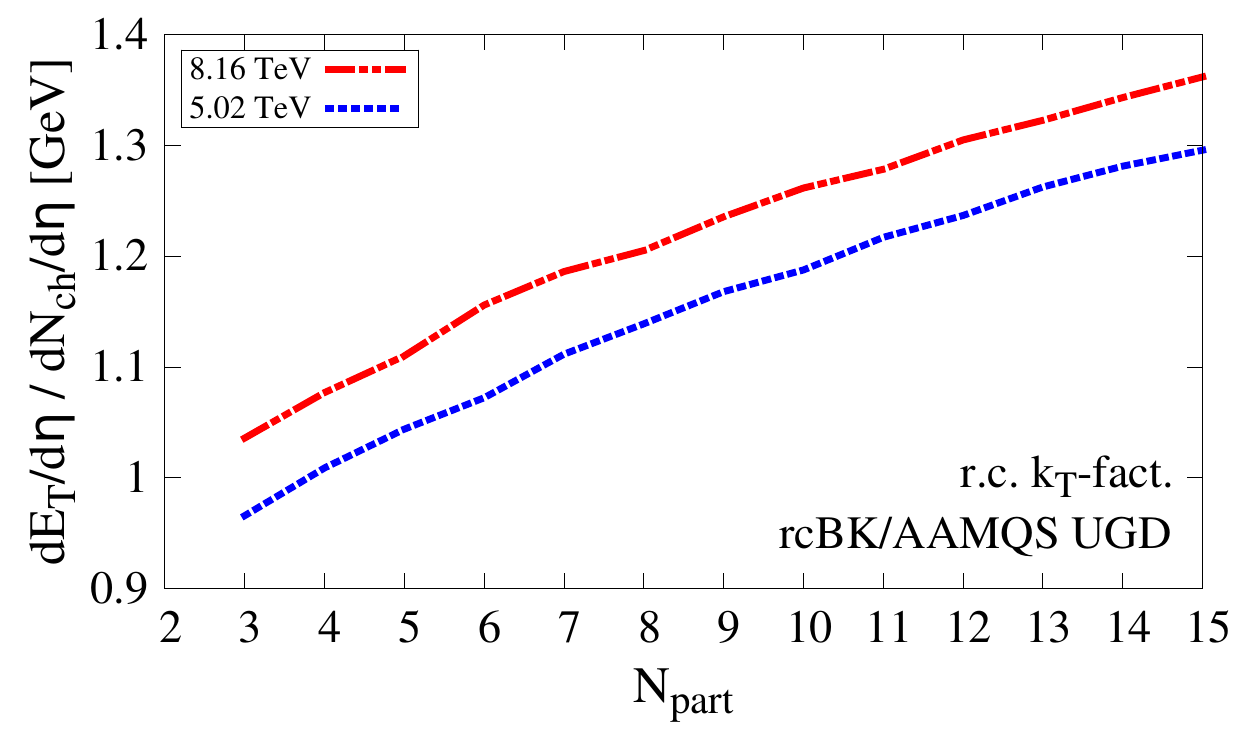}
\end{center}
\vspace*{-7mm}
\caption[a]{Transverse energy per charged particle vs.\ $N_\text{part}$
  in the central region of p+Pb collisions at 5~TeV and 8~TeV.}
\label{fig:pA_Et_Nch}
\end{figure}
\fig{fig:pA_Et_Nch} shows our result for the dependence of the
transverse energy divided by the number of charged particles, in p+Pb
collisions, on the number of participants. This ratio is independent of
the normalization factor $N$ in \eq{eq:rcktfact} and has instead been
normalized to the CMS measurement~\cite{CMS:2015opq} of $dE_T/d\eta$
in minimum bias p+Pb collisions at 5~TeV. The dependence on
$N_\text{part}$ and on energy is then a prediction. As before, the
number of participants should be determined in the fragmentation
region of the nucleus with a method that smoothly approaches the p+p
limit as $N_\text{part}\to 2$. This omits the bias on $E_T$ due to
fluctuations of the multiplicity at midrapidity which p+A collisions
inherit from p+p~\cite{Bzdak:2013lva}.

Our computation predicts an increase of $E_T$ per particle for
increasingly asymmetric collisions. Such ``broadening'' of the
transverse momentum distributions of produced gluons is expected due
to the increase of the saturation scale of the target
nucleus~\cite{Kharzeev:2001gp}. We should point out that $dE_T/d\eta$
is, however, sensitive to final state interactions which may reduce
its magnitude~\cite{dEt_red}.

One may attempt to interpret the decrease of the particle multiplicity per
participant with $N_\text{part}$ noted above in a simple two-component
``soft + hard'' model~\cite{Wang:2000bf}.  However we can see that
this is not possible.

Let $f(\sqrt{s})\ge0$ denote the fractional contribution from the hard
component which is proportional to the number of binary
collisions. $1-f(\sqrt{s})$ then corresponds to the soft contribution,
proportional to $N_\text{part}$:
\be \label{eq:soft+hard_AB}
\frac{dN_{AB}}{d\eta}= \left[\frac{1-f}{2}N_\text{part} + f
  N_\text{coll}\right] \frac{dN_{pp}}{d\eta}~.
\ee
In ref.~\cite{Wang:2000bf} Kharzeev and Nardi obtained
$f(\sqrt{s}=56~\text{GeV})=0.22$ and $f(\sqrt{s} = 130~\text{GeV}) =
0.37$. For $\sqrt{s}=5$~TeV we find that a universal fit from p+p to
central Pb+Pb with \eq{eq:soft+hard_AB} is impossible. Fitting to very
peripheral Pb+Pb collisions only ($N_\text{part}\le34$ corresponding
to $\le17$ participants per nucleus, on average) we estimate $f\approx
0.26\pm0.01$. A similar fit of the new Xe+Xe data by
ALICE~\cite{Acharya:2018hhy}, again for $N_\text{part}\le34$, gives
$f\approx 0.34\pm0.01$.  We consider this a lower bound on the value
of $f$ appropriate for p+A collisions since leading-twist perturbative
processes may already experience slight ``shadowing'' even in rather
peripheral heavy-ion collisions.

For p+A collisions we can rearrange the above equation as follows:
\be \label{eq:soft+hard_pA}
\frac{1}{N_\text{part}} \frac{dN_{pA}}{d\eta}= \left[\frac{1+f}{2} -
  \frac{f}{N_\text{part}}\right] \frac{dN_{pp}}{d\eta}~.
\ee
The r.h.s.\ is an increasing function of $N_\text{part}$ for any
$f>0$. The curve corresponding to the r.h.s.\ of \eq{eq:soft+hard_pA}
with $f=0.26\to 0.34$ and $dN_{pp}/d\eta=4.4$ is shown as a band in
\fig{fig:pA_roots_Npart}(right). The formula describes the data {\em
  below} the average $\langle N_\text{part}\rangle \simeq 8$ for
minimum bias collisions fairly well. However, the trend for more
central p+Pb collisions appears different from the data
shown. Furthermore, since $N_\text{coll}$ is linear in $N_\text{part}$
for p+A collisions, this simple model would not predict an increase of
the transverse energy per particle like in \fig{fig:pA_Et_Nch}.

Let us summarize the main points of this paper. We have performed the
first analysis of the energy and centrality dependence of particle
multiplicities in the central region of high energy p+A collisions
predicted by $k_T$-factorization with running coupling corrections,
and rcBK gluon distributions. We point out that the formula derived by
Horowitz and Kovchegov~\cite{Horowitz:2010yg} results in a
well-defined gluon transverse momentum
distribution\footnote{$k_T$-factorization, of course, does not
  correctly describe the gluon $p_T$-distribution below the saturation
  scale of the proton. However, the contribution from the region
  $p_T<Q_{s,p}$ to the $p_T$-integrated multiplicity is small when
  $Q_{s,p}\ll Q_{s,A}$.} down to $p_T\sim\Lmsbar$. Since this is
conceptually the lowest scale where a computation in perturbation
theory at running coupling level applies, this framework does not
require an ad-hoc cutoff on the transverse momentum spectrum of
produced gluons. A contribution from $p_T\sim\Lmsbar$ and below would
be genuinely non-perturbative.

Our numerical results show that the r.c.\ $k_T$-factorization formula
with rcBK gluon densities provides a good description of the energy
and centrality dependence of the multiplicity in p+A collisions at LHC
energies, $\sqrt{s}>1$~TeV. For p+A collisions at RHIC energies, on
the other hand, a better understanding of the unintegrated gluon
distribution of a proton at $x\sim0.01$ is required. It may be worth
pointing out that we have attempted to introduce as few model
parameters as reasonably possible in order to exhibit where the
current theory fails. In our analysis of particle multiplicities we
fitted a single energy, centrality, and system independent constant:
the normalization factor $N$ in \eq{eq:rcktfact}.

Our main observation relevant for phenomenology is to note that the
convolution of the unintegrated gluon densities of a proton and of a
nucleus increases more slowly than linear with the asymmetry of the
gluon densities set by the number of participants. The asymmetry of
the gluon distributions in p+A collisions results from the coherence
of the interaction with the dense target.  As a consequence, the
multiplicity per participant in increasingly asymmetric p+A collisions
is found to decrease slowly. (This may flatten out somewhat if a bias
on the small-$x$ gluon distribution is taken into account.) Such
behavior is markedly different from that for more and more central,
and increasingly symmetric A+A collisions, as well as from
expectations based on a simple two-component ``soft+hard'' particle
production model with only energy dependent shares.

\begin{acknowledgments}
We thank Yu.~Kovchegov, L.~McLerran and V.~Skokov for useful comments.
A.D.\ acknowledges support by the DOE Office of Nuclear
Physics through Grant No.\ DE-FG02-09ER41620; and from The City
University of New York through the PSC-CUNY Research grant 60262-0048.
A.V.G. thanks Sergio Korogui for helping with programming related
issues and gratefully acknowledges the Brazilian funding agency FAPESP
for financial support through grant 17/14974-8.  M.L.~acknowledges
support from FAPESP projects 2016/24029-6 and 2017/05685-2, and
project INCT-FNA Proc.~No.~464898/2014-5.
\end{acknowledgments}

\end{document}